\documentclass[pra,twocolumn,showpacs,preprintnumbers,amsmath,amssymb,superscriptaddress]{revtex4}

\usepackage{graphicx}
\usepackage{dcolumn}
\usepackage{bm}
\usepackage{color}
\usepackage{amsmath}
\usepackage{subfig}

\newcommand{\<}{\langle}
\renewcommand{\>}{\rangle}

\def\duzomniejsze{<\kern-.7mm<}
\def\duzowieksze{>\kern-.7mm>}

\def\textbf#1{{\bf #1}}
\def\beq{\begin{equation}}
\def\eeq{\end{equation}}
\def\be{\begin{equation}}
\def\ee{\end{equation}}
\def\ben{\begin{eqnarray}}
\def\een{\end{eqnarray}}
\def\beqa{\begin{eqnarray}}
\def\eeqa{\end{eqnarray}}
\def\eea{\end{array}}
\def\bea{\begin{array}}
\newcommand{\bei}{\begin{itemize}}
\newcommand{\eei}{\end{itemize}}
\newcommand{\bee}{\begin{enumerate}}
\newcommand{\eee}{\end{enumerate}}

\begin{document}


\title{Nonsignaling quantum random access code boxes}

\author{Andrzej Grudka}
\affiliation{Faculty of Physics, Adam Mickiewicz University, 61-614 Pozna\'n, Poland}

\author{Micha{\l} Horodecki}
\affiliation{Institute for Theoretical Physics and Astrophysics, University of Gda{\'n}sk, 80-952 Gda{\'n}sk, Poland}

\author{Ryszard Horodecki}
\affiliation{Institute for Theoretical Physics and Astrophysics, University of Gda{\'n}sk, 80-952 Gda{\'n}sk, Poland}

\author{Antoni W\'{o}jcik}
\affiliation{Faculty of Physics, Adam Mickiewicz University, 61-614 Pozna\'n, Poland}


\date{\today}

\begin{abstract}
A well known cryptographic primitive is so called {\it random access code}. Namely, Alice is to send to Bob one of two bits, so that Bob has the choice which bit he wants to learn about. However at any time Alice should not learn Bob's choice, and Bob should learn only the bit of his choice. The task is impossible to accomplish by means of either classical or quantum communication. 
On the other hand, a concept of correlations stronger than quantum ones, exhibited by so called  {\it Popescu-Rohrlich box}, was 
introduced and widely studied.  In particular, it is known that Popescu-Rohrlich box enables simulation of the random 
access code with the support of one bit of communication. Here, we propose a quantum analogue of this phenomenon. 
Namely, we define an analogue of a random access code, where instead of classical bits, one encodes qubits.
We provide a quantum non-signaling box that if supported with two classical bits, 
allows to simulate a quantum version of random access code. 
We point out  that two bits are necessary. We also show that a quantum random access code cannot be fully quantum:
when Bob inputs {\it superposition} of two choices, the output will be in a mixed state rather than in a superposition of required states.
\end{abstract}

\pacs{}
\maketitle
\section{Introduction} 
In recent years, we faced rapidly growing interest in analysing systems that obey the constraints of 
impossibility of instant transmission of messages. The constraints, called {\it non-signaling},  are satisfied by quantum mechanics, hence any limitations they pose
are also present in quantum mechanics. However, in so-called non-signaling theories, there are objects that exhibit behaviour 
forbidden by quantum mechanics. One of the basic blocks of non-signaling theories is the so called Popescu-Rohrlich box (PR-box) \cite{PR} -- a device that possesses much stronger correlations than those allowed by quantum mechanics.
It has a remarkable property of being able to simulate a {\it random access code} (RAC) with the support 
of only one bit of communication \cite{pawlowski-2009-461}.

Suppose that Alice wants to send to Bob one of two bits, so that Bob has the choice which bit he wants to learn about. Suppose further that the following conditions are met: first, when Bob gets perfect knowledge about bit, he must have no knowledge about the other bit, second, no communication from Bob to Alice is allowed, i.e., Bob should not tell Alice which bit he wants to learn, as well as after the 
execution of the protocol Alice should still not know which bit he learned.

Such a scenario is called random access code \cite{AmbainisNTV2002_qrac}.  
This task is impossible, when Alice and Bob share either classical or quantum states. 
However,  if Alice and Bob share the PR-box, they can implement it by sending just {\it one} bit. 
This peculiar feature  was used to formulate the principle of {\it information causality} \cite{pawlowski-2009-461}.  In this context 
in \cite{GrudkaHHKP-equiv-PR-rac} the notion of a {\it racbox} was introduced. It is a box which can implement a random access code with the support of one bit of communication. It was shown that any non-singaling racbox is equivalent to PR-box.

A natural question is whether one can have a quantum analogue of this phenomenon. 
Namely, we consider {\it quantum random access code}, where Alice has two qubits, and Bob wants to learn about the qubit of his choice.
Again, communication from Bob to Alice is not allowed, and  Bob should not  learn about the other qubit. 
Let us emphasize that this is a different concept from 
quantum random access code introduced in \cite{AmbainisNTV2002_qrac} and further considered in \cite{Pawlowski2010}
where qubits are used to simulate the standard random access code -- the one with classical inputs and outputs -- 
by encoding input classical bits into the quantum system, and then decoding the chosen classical bit by measurement.
In our case, both inputs and outputs of the quantum random access code are quantum states.

One can now ask, whether such functionality can be achieved by means of a {\it quantum non-signaling box}, \cite{PianiHHH-nosignaling-boxes} 
i.e.,  the non-signaling box that accepts qubits as inputs. Such a box can be viewed as a quantum channel, with two inputs and two outputs, with property, that the statistics of the output at one site do not depend on the input at the other site.
In this paper, we propose a quantum non-signaling box which, if supported by two bits of classical communication,
implements the above quantum version of RAC. 
The box is built out of two PR-boxes and two maximally entangled quantum states. 
We also prove that two bits of communication are necessary, by using analogy with quantum teleportation. We then show, that no  quantum non-signaling  box can give rise to a  fully quantum 
RAC. Namely,  if Bob inputs {\it superposition} of decisions on which qubit he wants to learn about,
the output must be a mixture of states of Alice's qubits rather than superposition.
This resembles the question of whether quantum computer can be fully quantum  asked in \cite{LindenP-halting}, where the superposition of halt times was impossible. 

\section{Random access code and ``racbox''}

In this section we describe the standard random access code, and a closely related object called ``racbox''. Namely, suppose that Alice has two bits $x_0$ and $x_1$, 
and Bob wants to learn one of them.  We want Bob to have a choice, which bit he would like to learn, but if he learns 
one of the bits, then the other should be lost. Moreover at any time Alice cannot know Bob's choice. As already mentioned in introduction, such a task is called random access code.

There does not exist classical or quantum communication protocol, 
that can perform this task, which is easy to see at least in classical case. Indeed, 
the only thing Alice can do, so that Bob can read the bit of his choice, is to send both bits. But in such case the condition 
that he should not learn the other bit is not met. Let us also note, that if we weaken the definition 
of random access code and  will not assume, that Bob cannot learn two bits, then such a weaker version of random access code 
needs two bits of classical communication. 

\begin{figure}
\includegraphics[trim= 8cm 4cm 8cm 1cm, clip=true, width=8cm]{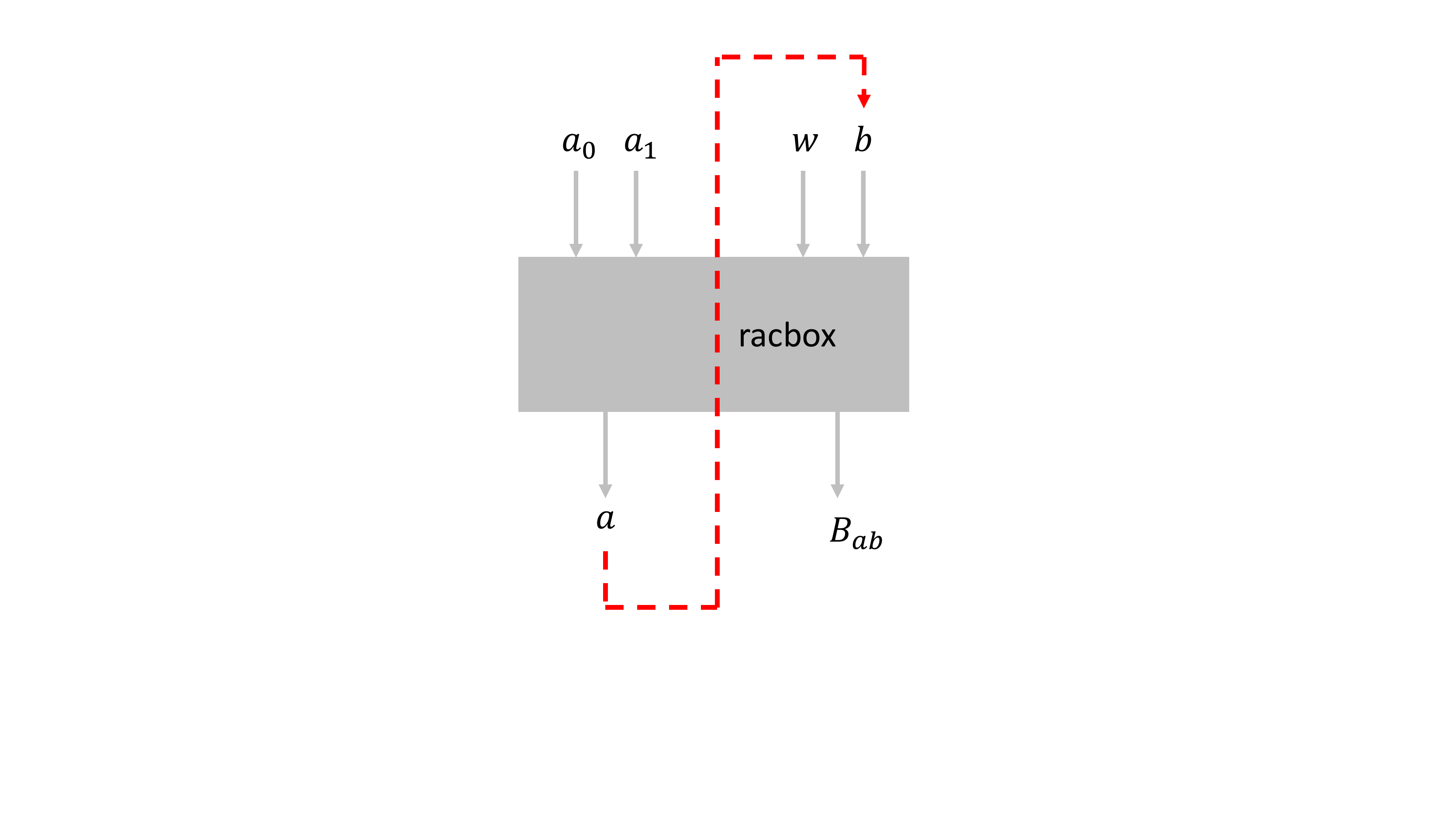}
\caption{Racbox. Alice has two binary inputs $a_0$ and $a_1$, and binary output $a$. 
Bob has two binary inputs $b,w$ and binary output $B_{ab}$. When $b=a$ Bob's output is equal to $a_w$, i.e. it depends on Bob's input $w$. Hence, if Alice sends her output  to Bob, he can read Alice's bit of his choice. \label{fig:racbox}}
\end{figure}

The situation changes if Alice and Bob  share so called PR-box. PR-box is a bipartite device shared by two distant parties Alice and Bob.
Each of the parties can choose one of two inputs: Alice $x=0,1$ and Bob $y=0,1$.
The parties have two binary outputs $a,b$. The box is defined by a family of joint probability distributions $p(ab|xy)$ which satisfy
\be
p(ab|xy)=
\left\{
\bea{l}
\frac12 \quad{\textrm{for }} a\oplus b=xy, \\
0 \quad{\textrm{else}}.
\eea
\right.
\ee
The PR box can be interpreted in two ways. On one hand it can be considered as a ``super-quantum'' resource, as it allows for correlations,
that cannot be obtained from measuring bipartite quantum state. On the other hand, 
it can be treated as a  classical channel, with two remote inputs and two remote outputs. The channel has a special property:
its implementation requires 1 bit of communication, but  if it works as a ``black box'' - 
i.e. if the parties can only use the box through the inputs and outputs, it cannot be used for communication - we say it is 
non-signaling. Now, in \cite{WW} it is shown, that if Alice and Bob share a PR box, they can implement random access code 
by means of just one bit of communication. In \cite{GrudkaHHKP-equiv-PR-rac} a converse question was answered: 
Namely,  an object was defined called {\it racbox}. It is a box that implements RAC if supported by one bit of communication from Alice to Bob (see 
Fig. \ref{fig:racbox}).
It was then shown that a {\it non-sinaling} racbox is equivalent to PR box.

\section{QRAC-box} 
\begin{figure}
{\includegraphics[trim= 8cm 4cm 8cm 2cm, clip=true, width=8cm]{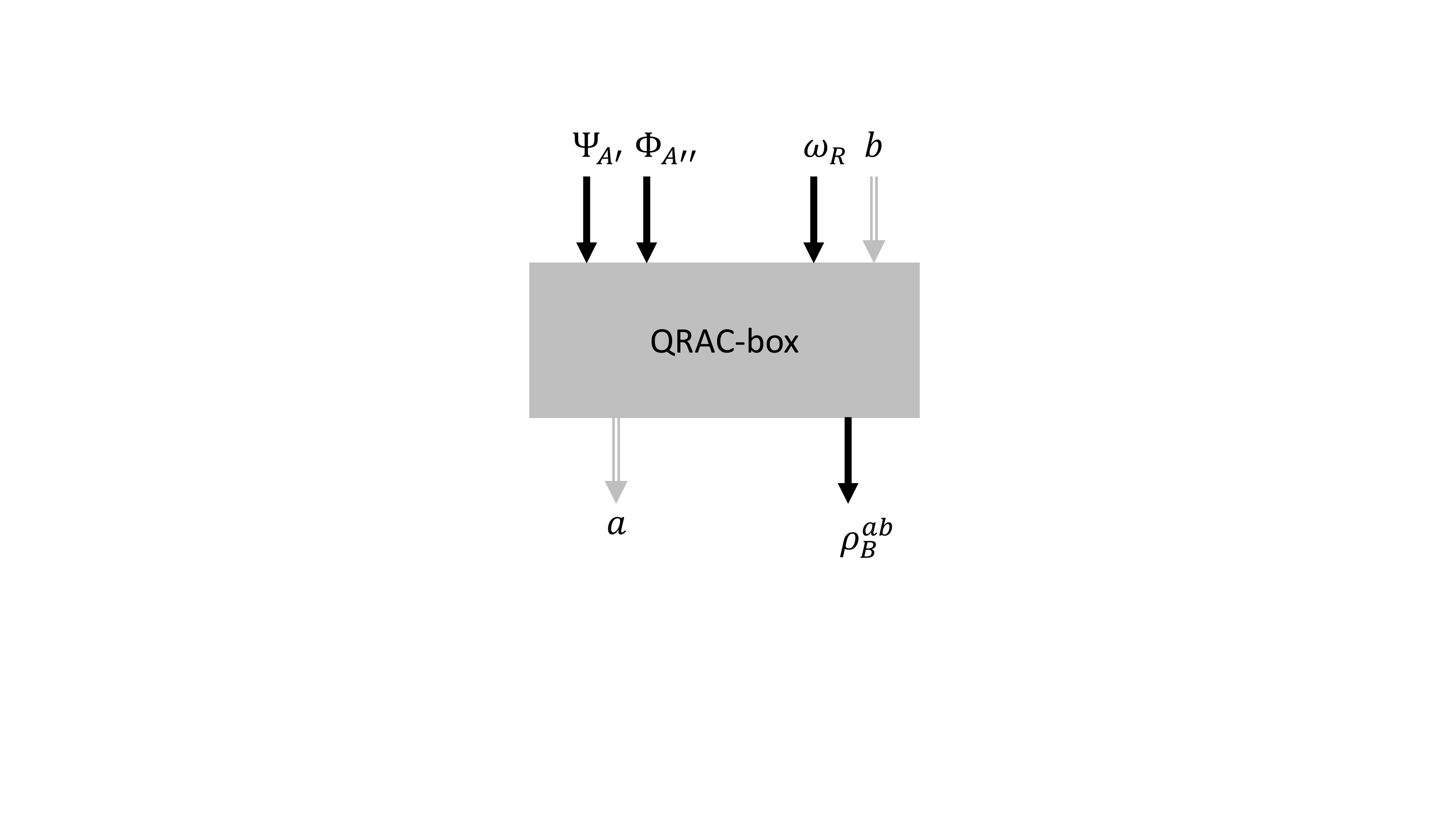}}
\caption{QRAC-box. Alice has two qubit inputs $\Psi_{A'}$ and $\Phi_{A''}$, and two-bit classical output $a$. Bob has one qubit input $\omega_R$, two-bit classical input $b$, and qubit output $\rho_B^{ab}$. When $b=a$ then depending on Bob's input $\omega_R$ his output $\rho_B^{ab}$ is equal to $\Psi_{A'}$ or $\Phi_{A''}$. \label{fig:qrac}}
\end{figure}

In this section we define non-signaling quantum random access code box (QRAC-box, cf. \cite{GrudkaHHKP-equiv-PR-rac}), which performs a quantum version of random access code if supplemented with 2 bits of communication. QRAC-box is a bipartite device shared by Alice and Bob.
Alice has a two-qubit input and a two-bit classical output (later we show that this is the smallest possible size of Alice's classical output. Bob has two inputs: a one-qubit input and a two-bit classical input. He also has a one-qubit output (see Fig. \ref{fig:qrac}). 

We assume that the device obeys quantum mechanical laws, i.e., it is trace preserving 
completely positive map. We further assume that device cannot signal from one party to the other party, i.e., 
one party's output cannot depend on the other party's input.  
Now, such a  device will be called QRAC-box, if it possesses the following property. 
Suppose that Alice inputs the first qubit in a state $|\Psi\>_{A'}$ and the second qubit in a state $|\Phi\>_{A''}$. She then obtains $a$ as her output. When Bob's classical input $b$ is equal to Alice's classical output $a$ and his input qubit is in a state $|0\>_R$ then we require that he obtains a state $|\Psi\>_B$ as his output. On the other hand, when Bob's input $b$ is equal to Alice's output $a$ and his input qubit is in a state $|1\>_R$, then we require that he obtains a state $|\Phi\>_B$ as his output. 
As a result, if Alice sends her output to Bob, then Bob can obtain Alice's qubit of his choice, 
by simply inputing $b=a$. 

Let us note, that from the fact that device is non-signaling, i.e., in particular,
Alice's output does not depend on Bob's input, the above definition of QRAC-box is consistent, i.e., the classical output 
of Alice can be fed as Bob's input without causing a contradiction.  If, on the contrary, the output of Alice would depend on input of Bob,
then it might happen that whenever Bob wants to input $b=a$, then this changed output of Alice, 
so that it were no longer $a$ and we would obtain a contradiction, i.e., Bob would not be able to input Alice's output.

\begin{figure}
{\includegraphics[trim= 3cm 2cm 3cm 2cm, clip=true, width=8cm]{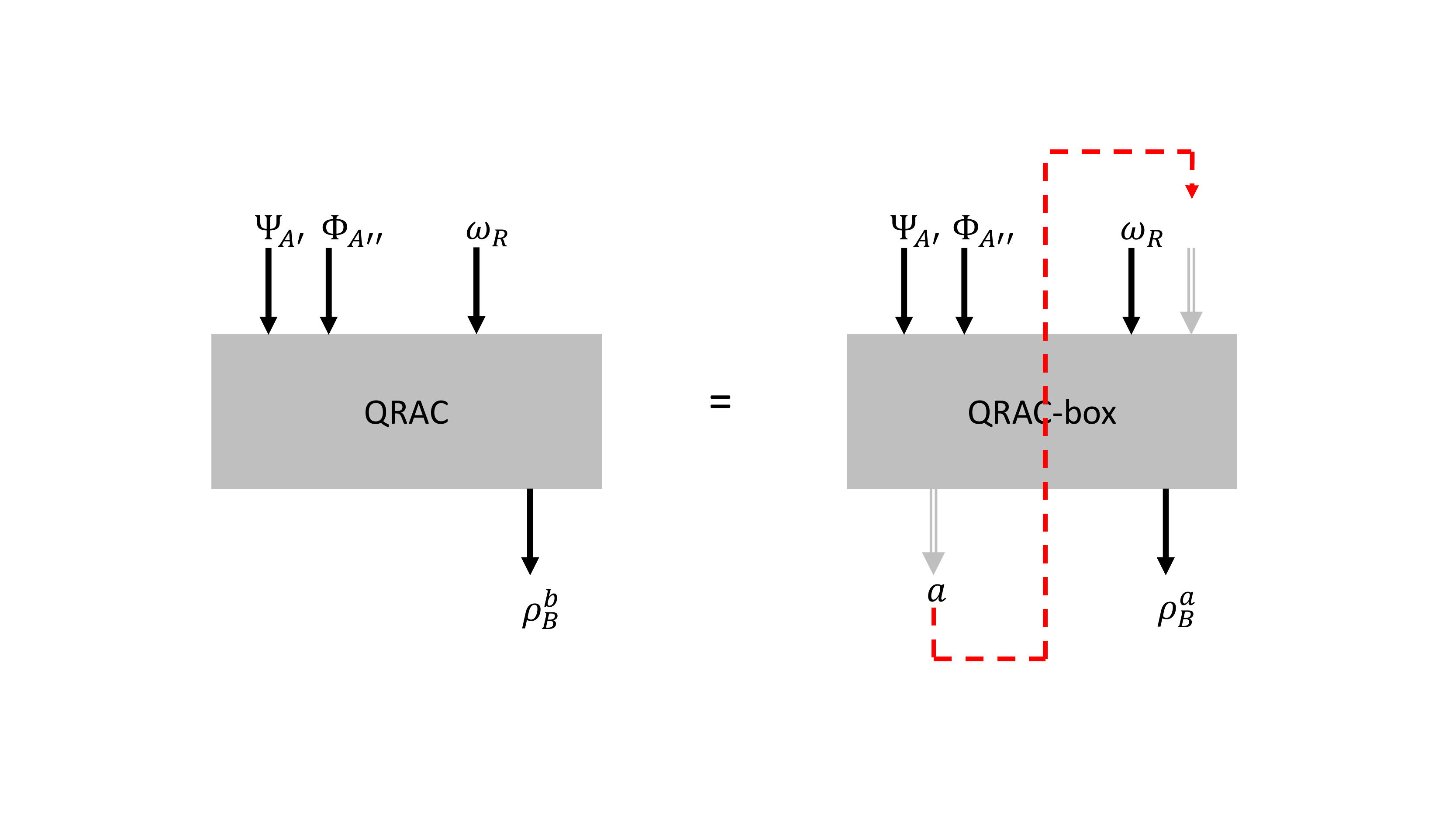}}
\caption{The channel QRAC resulting from feeding Alice's classical  output to Bob's classical input of QRAC-box. \label{fig:lambda}}
\end{figure}

Finally, let us note, that the above properties of QRAC-box imply, that the box obtained from feeding Alice's classical 
input as Bob's classical input is a quantum channel  too (which may not be no-signaling anymore) 
with three inputs, and one output, see Fig. \ref{fig:lambda}. We shall call the channel QRAC. 
It is a sum of subchannels (which are completely positive trace non-increasing maps) labeled by Alice's outputs
\be
\Lambda=\sum_a\Lambda_a.
\label{eq:lambda}
\ee
where by $\Lambda$ we denote the channel representing QRAC.

As mentioned above, we assumed  that QRAC-box obeys the laws of quantum mechanics, i.e.,  it is a quantum channel (with several inputs and outputs) of some particular features. 
A possible way to implement such a channel in lab is to send inputs from Alice and Bob to a joint place,
perform the quantum operation, e.g., by means of a circuit composed of quantum gates (nowadays more and more complicated circuits 
are possible to implement in labs), and resend the outputs of the channel back to Alice and Bob. 
In such a scenario, to implement the channel, quantum communication is required. However, from  the point of view of Alice and Bob,
our channel is a black box. Hence, it cannot be used to signal from Alice to Bob and vice versa.  
Thus the situation is analogous to the case of the PR box - the latter is a classical channel, and one can implement it by means of classical communication, yet considered as a black box, it cannot be used itself to perform communication.

One can also consider another way of implementing such channels through pre- and post-selection 
as proposed in \cite{PhysRevA.75.022102} and realised experimentally in  \cite{PhysRevA.80.030101}. 
The two ways, are strictly connected. Since the channel requires communication to implement it, 
if one wants to implement it without communication, one needs to consider some pre- or post-selection
(which is a hidden form of communication).

\section{Superposition} 
Let us suppose that instead of preparing his qubit in a state $|0\>_R$ or $|1\>_R$ and decoding the first or the second of Alice's qubits Bob prepares his qubit in a state $|\omega\>_R=\alpha|0\>_R+\beta|1\>_R$. What will his output state be when his classical input is equal to Alice's output? Will he obtain a superposition of states $|\Psi\>_B$ and $|\Phi\>_B$? Below we answer these questions.

First we will show that the channel QRAC defined in the previous section produces a mixture of those states, rather than superposition. 
Then we will argue, that each of subchannels $\Lambda_a$ also produces such a mixture (now subnormalized).


Consider then $\Lambda$ of Eq. \ref{eq:lambda}. We extend this trace preserving completely positive map to unitary operation $U$ acting on a system and environment. Let us check how it acts when Bob prepares his qubit in a state $|0\>_R$ and $|1\>_R$ and his input $b$ is equal to Alice's output $a$.  We have
\ben
&&U(|\Psi\>_{A'}\otimes|\Phi\>_{A''}\otimes|0\>_R\otimes|\chi\>_E)=|\Psi\>_B\otimes|\chi^{(0)}\>_{A''RE} \nonumber \\  
&&U(|\Psi\>_{A'}|\otimes|\Phi\>_{A''}\otimes|1\>_R\otimes|\chi\>_E)=|\Phi\>_B\otimes|\chi^{(1)}\>_{A''RE}
\een
where we renamed Alice's first input register as Bob's output register, $|\chi\>_E$ is the initial state of the environment while $|\chi^{(0)}\>_{A''RE}$ and $|\chi^{(1)}\>_{A''RE}$ are the final states of Alice's second register, Bob's input register $R$ and environment.
Let us note that the states $|\Psi\>_{A'}\otimes|\Phi\>_{A''}\otimes|0\>_R\otimes|\chi\>_{E}$ and $|\Psi\>_{A'}\otimes|\Phi\>_{A''}\otimes|1\>_R\otimes|\chi\>_{E}$ are orthogonal. Hence either $|\Psi\>_{B}$ is orthogonal to $|\Phi\>_{B}$, or $|\chi^{(0)}\>_{A''RE}$ is orthogonal to $|\chi^{(1)}\>_{A''RE}$. Since for all $|\Psi\>_{B}$ non-orthogonal to $|\Phi\>_{B}$, the state  $|\chi^{(0)}\>_{A''RE}$ is orthogonal to 
$|\chi^{(1)}\>_{A''RE}$, then by continuity for $|\Psi\>_{B}$ orthogonal to $|\Phi\>_{B}$ the state $|\chi^{(0)}\>_{A''RE}$ has to be orthogonal to $|\chi^{(1)}\>_{A''RE}$ as well.
When Bob prepares his qubit in a state $\alpha|0\>_R+\beta|1\>_R$ then by linearity we have
\begin{eqnarray}
&& U|\Psi\>_{A'}\otimes|\Phi\>_{A''}\otimes(a|0\>_R+b|1\>_R)\otimes|\chi\>_E=\nonumber \\
&& \alpha|\Psi\>_B\otimes|\chi^{(0)}\>_{A''RE}+\beta |\Phi\>_B\otimes|\chi^{(1)}\>_{A''RE}.
\end{eqnarray}
Tracing out Alice's second register, Bob's input register and environment and using orthogonality of states $|\chi^{(0)}\>_{A''RE}$ 
and $|\chi^{(1)}\>_{A''RE}$ we obtain that Bob's output state is
\be
\rho_{B}=|\alpha|^2|\Psi\>\<\Psi|_{B}+|\beta|^2|\Phi\>\<\Phi|_{B}.
\label{eq:mixture}
\ee
We see that Bob obtains a mixture rather than a superposition of states  $|\Psi\>_B$ and $|\Phi\>_B$.

Now, consider the subchannels $\Lambda_a$, and suppose, by contradiction, 
that some of them produces a state which is not equal to such mixture. Let us denote outputs by $\rho_B^a$.
One then easily sees, that there exist unitaries $U_a$ such that 
\be
\sum_a U_a \rho_B^a U_a^\dagger \not =  |\alpha|^2|\Psi\>\<\Psi|_{B}+|\beta|^2|\Phi\>\<\Phi|_{B}
\label{eq:non_mixture} 
\ee
Thus, we consider QRAC-box with the above subchannels $\Lambda_a$, and we will construct a new QRAC-box
as follows: Bob while inputting $b$ will apply transformation $U_b$  to his output. 
One checks that it defines a valid QRAC-box, resulting in  subchannels $\Lambda'_a = \hat U_a \Lambda_a$,
where $\hat U_a( \cdot) = U_a (\cdot) U_a^\dagger$. Therefore, due to \eqref{eq:non_mixture} the resulting QRAC given by 
$\Lambda'=\sum_a \Lambda'_a$ will produce a state which is not a mixture  \eqref{eq:mixture}. 
However this contradicts to our first result, that QRAC resulting from arbitrary QRAC-box, necessarily produces mixture \eqref{eq:mixture}.

\section{Communication cost} 
Let us now find lower bound for minimal amount of classical information which Alice has to send to Bob so that he can retrieve Alice's qubit of his choice. In the next section we present a box which achieves this bound. Let us assume that Bob prepares his input qubit in a state $|0\>_R$ and tries to obtain Alice's first qubit (similar analysis applies when Bob prepares his input qubit in a state $|1\>_R$ and tries to obtain  Alice's second qubit). We  know from the previous section that there is no need to consider the case when Bob prepares his qubit in a state $\alpha|0\>_R+\beta|1\>_R$ as he can simply measure it and depending on a result of the measurement input a state $|0\>_R$ or $|1\>_R$. In the case when Bob prepares his input qubit in a state $|0\>_R$ QRAC-box acts just like quantum teleportation.
Indeed, if Alice sends her classical output to Bob and Bob uses it as his classical input, then he obtains Alice's first qubit.  
Now, since we require that QRAC-box is non-signaling, we just need to argue, that if Alice and Bob have non-signaling resources, then they need at least two bits to perform teleportation. 
However, this was already proven in \cite{Teleportation33}. Namely it is argued that by combining quantum teleportation with dense coding, one would obtain instantaneous communication, thereby violating causality.

\begin{figure}
{\includegraphics[trim= 0cm 1cm 1cm 3cm, clip=true, width=12truecm]{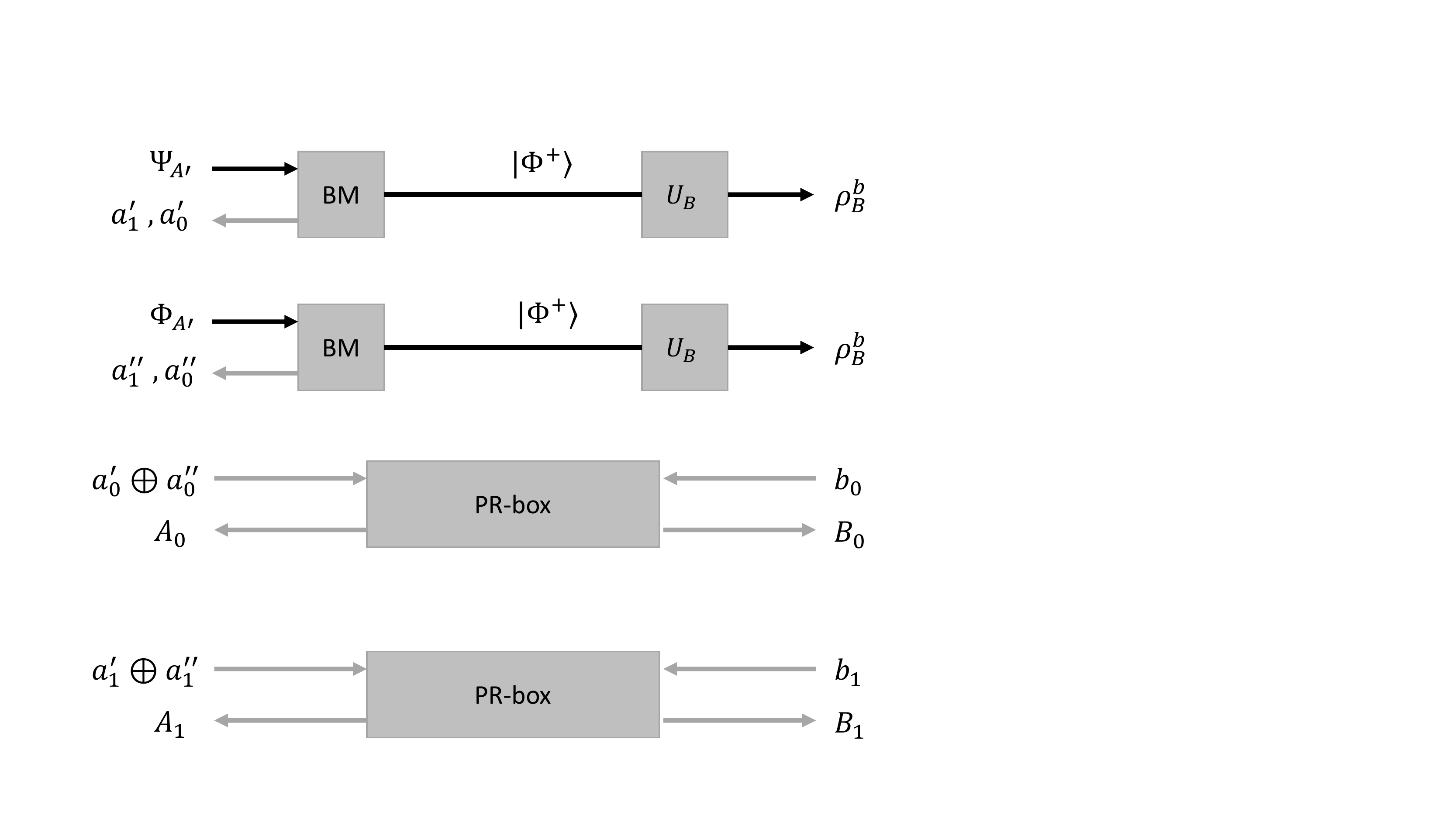}}
\caption{Implementation of QRAC-box with entanglement and PR-boxes. Alice and Bob share two maximally entangled pairs and two PR-boxes. Alice performs two Bell measurements (BM) -- the first one on her first input qubit and her qubit from the first maximally entangled pair, and the second one on her second input qubit and her qubit from the second maximally entangled pair. Results of the measurements are represented by two two-bit strings -- $a'_1a'_0$ in the case of the first measurement and $a''_1a''_0$ in the case of the second measurement. She inputs $a'_0 \oplus a''_0$ into the first PR-box and $a'_1 \oplus a''_1$ into the second PR-box and obtains outputs $A_0$ and $A_1$ respectively. Alice's classical two-bit output $a=a_1a_0$ (see Fig. \ref{fig:qrac}) is given by $a_0=a'_0\oplus A_0$ and $a_1=a''_0 \oplus A_1$. If Bob wants to obtain the first (second) Alice's qubit he inputs $0$ ($1$) into both PR-boxes and obtains outputs $B_0$ and $B_1$. Then he applies one of four unitary operations ($U_B$) to his qubit from the first (second) maximally entangled pair and discards the other qubit. The choice of unitary operation depends on $B_0\oplus a_0$ and $B_1\oplus a_1$. \label{fig:EPRPR}}
\end{figure}

\section{Implementation of QRAC-box with entanglement and PR-boxes} 
We show how one can simulate QRAC-box with two maximally entangled pairs and two PR-boxes. The protocol is based on quantum teleportation and implementation of classical RAC with PR-boxes  (see Fig. \ref{fig:EPRPR}).
Let us suppose that Alice and Bob apart from qubits which they input into the box share two pairs of qubits and two PR-boxes. Each pair of qubits is in the maximally entangled state
\begin{eqnarray}
|\Phi^{+}\rangle_{AB}=\frac{1}{\sqrt{2}}(|0\rangle_{A}\otimes|0\rangle_{B}+|1\rangle_{A}\otimes|1\rangle_{B})
\end{eqnarray}
PR-box has two inputs and two outputs -- one input and output is on Alice's side and one input and output is on Bob's side. When Alice and Bob input $a$ and $b$ into the box they obtain outputs $A$ and $B$ with probability
\begin{eqnarray}
& p(AB|ab)=\frac{1}{2} \text{ if } A \oplus B=ab \nonumber \\
& p(AB|ab)=0 \text{ otherwise }
\label{PR-box}
\end{eqnarray}
In order to implement QRAC Alice performs measurement in the Bell basis
\be
X_{A'}^{a'_0}Z_{A}^{a'_1}|\Phi^{+}\rangle_{A'A} \hspace{0.5cm} (a'_0,a'_1 \in \{0,1\})
\ee
on the first qubit $|\Psi\rangle_{A'}$ and her qubit from the first maximally entangled pair and obtains two-bit result $a'_1a'_0$. After the measurement Bob's qubit from the first maximally entangled pair is in the state $X_B^{a'_0}Z_B^{a'_1}|\Psi\rangle_{B}$. Similarly, Alice performs measurement in the Bell basis on the second qubit $|\Phi\rangle_{A'}$and her qubit from the second maximally entangled pair and obtains two-bit result $a''_1a''_0$. Now Alice inputs $a'_0 \oplus a''_0$ into the first PR-box, and $a'_1 \oplus a''_1$ into the second PR-box and obtains outputs $A_0$ and $A_1$. Alice's output bits of the QRAC-box will be $a_0=a'_0 \oplus A_0$ and $a_1=a'_1 \oplus A_1$. Next Alice sends two-bit message $a_1a_0$ to Bob. If Bob wants to obtain Alice's first qubit (corresponding to the state of his qubit input $|0\rangle_R$) he inputs $0$ both into the first PR-box and into the second PR-box. He obtains outputs $B_0$ and $B_1$ respectively. He then calculates $b_0=a_0\oplus B_0= a'_0\oplus A_0 \oplus B_0=a'_0$ and $b_1=a_1\oplus B_1=a'_1 \oplus A_1 \oplus B_1=a'_1$. The last equality in each expression follows from Eq. \ref{PR-box}. Finally he applies unitary operation $Z_B^{b_1}X_B^{b_0}$ to his qubit from the first maximally entangled pair. If Bob wants to obtain Alice's second qubit (corresponding to state of his qubit input $|1\rangle_R$) he inputs $1$ into both the first PR-box and the second PR-box, obtains outputs $B_0$ and $B_1$, calculates $b_0$ and $b_1$ (now $b_0= a'_0\oplus A_0 \oplus B_0=a'_0\oplus a'_0 \oplus a''_0=a''_0$ and $b_1=a'_1 \oplus A_1 \oplus B_1=a'_1\oplus a'_1 \oplus a''_1=a''_1$) and applies unitary operation $Z_B^{b_1}X_B^{b_0}$ to his qubit from the second maximally entangled pair. After application of unitary operation the qubit will be in a state equal to the initial state of the first (second) of Alice's qubits. Bob also discards his qubit from the second (first) maximally entangled pair.
In a general case when Bob prepared his qubit input in a state $\alpha|0\>_R+\beta|1\>_R$ he first performs a measurement on it in computational basis and then conditioned on the result of the measurement he decodes one of Alice's qubits. Let us note, that the constructed 
box is non-signaling, since it is obtained by local operations on non-signaling resources such as PR boxes and maximally entangled states.
Also our construction satisfies the condition that given Bob's classical input and Alice's classical output 
the transformation from Alice's input quantum state to Bob's output quantum state is a trace preserving completely positive map. Indeed, 
the transformation results from some local quantum operations and classical communication --  where communication is used to 
implement PR boxes.

We also note, that by applying dense coding, we can change the proposed QRAC-box into one that operates solely 
with qubits, i.e., instead of Alice's two-bit output, and Bob's two-bit output, they will have 1 qubit output and input, respectively.
In more detail,  our "qubit-only" QRAC-box will consist of the original QRAC-box, supplemented by maximally entangled pair. 
The two bits of outputs will be sent by means of this pair. Note that the pair will be treated as a part of the qubit-only QRAC-box, 
and will not be seen by users of the box, who will only see inputs and outputs, now all of them quantum. 
Thus, we obtain that a quantum random access code can be performed 
by use of a quantum non-signaling box supplemented by one qubit of communication.

\section{Summary} 
We introduced a non-signaling quantum random access code box -- a device which enables Bob to obtain one of two of Alice's qubits when Alice sends Bob two bits of classical information. It is important that Bob can choose which qubit he wants to obtain. We investigated properties of such a box and showed that two bits is minimum amount of classical information which Alice has to send to Bob, i.e., if there was less communication, the box must be signaling. We also showed how the box can be implemented with entanglement and PR-boxes.

\acknowledgements
We thank W. K{\l}obus and M. Piani for valuable comments on the manuscript.
This work is supported by the ERC Advanced Grant QOLAPS, The National Centre for Research and Development Grant QUASAR and National Science
Centre project Maestro DEC-2011/02/A/ST2/00305. Part of this work was done in National Quantum Information Centre of Gdansk (KCIK).

\end{document}